\documentclass[fleqn,usenatbib]{mnras}

\usepackage{newtxtext,newtxmath}
\usepackage{xcolor}
\usepackage{enumerate}
\usepackage[T1]{fontenc}

\DeclareRobustCommand{\VAN}[3]{#2}
\let\VANthebibliography\thebibliography
\def\thebibliography{\DeclareRobustCommand{\VAN}[3]{##3}\VANthebibliography}

\usepackage{graphicx}	
\usepackage{amsmath}	

\usepackage{amssymb}	
\usepackage{bm}
\usepackage{amsfonts}
\usepackage{latexsym}
\usepackage[latin1]{inputenc}
\usepackage{amsmath}
\usepackage{epsfig}
\usepackage{amsbsy}

\newcommand{\Reply}[1]{\textcolor{black}{#1}}

\title[Quasi-periodic oscillations]{Quasi-periodic oscillations during magnetar giant flares in the strangeon star model}

\author[H.-B. Li et al.]{
Hong-Bo Li,$^{1,2}$
Yacheng Kang,$^{1,2}$
Zexin Hu,$^{1,2}$ Lijing Shao,$^{2,3}$\thanks{E-mail: lshao@pku.edu.cn (LS)} 
Cheng-Jun Xia,$^{4}$
and Ren-Xin Xu$^{1,2}$\thanks{E-mail: r.x.xu@pku.edu.cn (R-XX)}
\\
$^{1}$Department of Astronomy, School of Physics, Peking University, Beijing 100871, China\\
$^{2}$Kavli Institute for Astronomy and Astrophysics, Peking University, Beijing 100871, China\\
$^{3}$National Astronomical Observatories, Chinese Academy of Sciences, Beijing 100012, China \\
$^{4}$Center for Gravitation and Cosmology, College of Physical Science and Technology, Yangzhou University, Yangzhou 225009, China
}

\date{Accepted XXX. Received YYY; in original form ZZZ}

\pubyear{2023}

\begin{document}
\label{firstpage}
\pagerange{\pageref{firstpage}--\pageref{lastpage}}
\maketitle

\begin{abstract}
Soft gamma-ray repeaters (SGRs) are widely understood as slowly rotating
isolated neutron stars. Their generally large spin-down rates, high magnetic
fields, and strong outburst energies render them different from ordinary
pulsars. In a few giant flares (GFs) and short bursts of SGRs,  high-confidence
quasi-periodic oscillations (QPOs) were observed. Although remaining an open
question, many theoretical studies suggest that the torsional oscillations
caused by starquakes could explain QPOs. Motivated by this scenario, we
systematically investigate torsional oscillation frequencies based on the
strangeon-star (SS) model with various values of harmonic indices and overtones.
To characterize the strong-repulsive interaction at short distances and the
non-relativistic nature of strangeons, a phenomenological Lennard-Jones model is
adopted. We show that, attributing to the large shear modulus of SSs, our
results explain well the high-frequency QPOs ($\gtrsim 150$\,Hz) during the GFs.
The low-frequency QPOs ($\lesssim 150$\,Hz) can also be interpreted when the
ocean-crust interface modes are included. We also discuss possible effects of
the magnetic field on the torsional mode frequencies. Considering realistic
models with general-relativistic corrections and  magnetic fields, we further
calculate torsional oscillation frequencies for quark stars. We show that it
would be difficult for quark stars to explain all QPOs in GFs. Our work advances
the understanding of the nature of QPOs and magnetar asteroseismology.
\end{abstract}


\section{Introduction}\label{sec:Introduction}

Magnetars are the most highly magnetized neutron stars (NSs) in the Universe
with typical magnetic fields $B \gtrsim 10^{14}$\,G. Judging from burst
activities and other aspects,  magnetars are classified into accretion-driven
X-ray pulsars, X-ray bursters, anomalous X-ray pulsars (AXPs), soft gamma-ray
repeaters (SGRs), etc. \citep{Coelho:2012jy, Turolla:2015mwa}. Among them,
AXPs/SGRs are widely known as slowly rotating isolated NSs. Compared with
ordinary pulsars, AXPs/SGRs usually have larger spin-down rates of $\dot{P} \sim
10^{-13}$--$10^{-10} \, \rm s \,s^{-1}$ with rotational periods in a narrower
range of $P \sim 2$--$12$\,s. In addition to persistent emissions and short
bursts,  giant flares (GFs) were observed associated with some SGRs. The peak
luminosity during such GFs is $10^{6}$ times greater than the Eddington
luminosity of a typical NS \citep{Mereghetti:2008je, Huang:2014mna}.  These
giant flares are typically accompanied by a decaying tail that lasts several
hundred seconds.  Some recorded GFs are as follows: SGR 0526$-$66 in 1979
\citep{Mazets:1979, Barat:1983}, SGR 1900+14 in 1998 \citep{Hurley:1998ks}, SGR
1806$-$20 in 2004 \citep{Terasawa:2005xg, Palmer:2005mi}, SGR J1550$-$5418 in
2014 \citep{Huppenkothen:2014pba}, SGR J1935+2154 in 2022 \citep{Li:2022jxj},
and SGR 150228213 in 2023 \citep{Chen:2023xnf}.

With different timing analysis methods, many studies have revealed the existence
of characteristic quasi-periodic oscillations (QPOs) during these GFs. For
different SGRs, the recorded QPOs are as follows: $43.5$\,Hz for SGR 0526$-$66
\citep{Barat:1983}; $18\pm 2$\,Hz, $26 \pm 3$\,Hz, $30 \pm 4$\,Hz, $92 \pm
2$\,Hz, $150 \pm 17$\,Hz, $625 \pm 2$\,Hz, and $1837 \pm 5$\,Hz for SGR
1806$-$20 \citep{Israel:2005av, Watts:2005ue, Strohmayer:2006py}; $28 \pm
2$\,Hz, $53 \pm 5$\,Hz, $84$\,Hz, and $155 \pm 6$\,Hz for SGR1900+14
\citep{Strohmayer:2005ks}; $93 \pm 12$\,Hz, $127 \pm 10$\,Hz, and possibly
$260$\,Hz for SGR J1550$-$5418 \citep{Huppenkothen:2014pba}; $40$\,Hz for SGR
J1935+2154 \citep{Li:2022jxj}; $60$\,Hz and $110$\,Hz for SGR 150228213
\citep{Chen:2023xnf}. QPOs are also found in other phenomena. Some QPOs have
been detected in gamma-ray bursts (GRBs): $22$\,Hz and $51$\,Hz for GRB 211211A
\citep{Xiao:2022quv}; $836$\,Hz, $1444$\,Hz, $2132$\,Hz, and $4250$\,Hz for GRB
200415A \citep{Castro-Tirado:2021}; $0.0015$\,Hz for GRB 180620A
\citep{Zou:2022rgf}. Some QPOs are found in fast radio bursts (FRBs): $4.6$\,Hz
for FRB 20191221A \citep{CHIMEFRB:2021fvq}; $2409$\,Hz for FRB 20201020A;
$93.46$\,Hz for FRB 20210213A; $357.14$\,Hz for FRB 20210206A; $1052.63$\,Hz for
FRB 20180916B A17; $588.24$\,Hz for FRB 20180916B A53
\citep{Pastor-Marazuela:2022pnp}.

Although the origin of QPOs remains uncertain, there are many theoretical
studies on the nature of such events. Based on the ordinary NS scenarios, some
models explore the idea that QPOs in some GFs are produced by the torsional
oscillations of the solid crust alone or global seismic vibration modes
\citep{Duncan:1998my, Piro:2005jf, Strohmayer:2005ks, Samuelsson:2006tt,
Sotani:2006at}. However, not all observed frequencies can be explained by these
attempts. Considering the non-negligible effects of the high magnetic fields,
superfluid, and nuclear pasta phases, there have been many investigations on the
global oscillations of magnetars. \citep[see][and references
therein]{Glampedakis:2006apa, Levin:2006ck, Levin:2006qd, Cerda-Duran:2009hdu,
Colaiuda:2010pc, Colaiuda:2011aa, Colaiuda:2009ne, Gabler:2010rp, Gabler:2011am,
Gabler:2013ova, Gabler:2012jh, Gabler:2016rth, Gabler:2017lvk, vanHoven:2010gy,
vanHoven:2011it, Passamonti:2013zra}. 

Alternatively, involving strange quark may shed light on the physical mechanism
of the QPOs in GFs. 
\Reply{ 
It is conjectured that the
bulk dense matter may be composed of strangeons, which are formerly named strange-quark clusters with
nearly equal numbers of $u$, $d$, and $s$ quarks \citep{Xu:2003xe}.
Based on phenomenological analysis and comparison with different observations, a strangeon star (SS) model was proposed with very stiff equation of states~\citep[EOSs; ][]{Xu:2003xe, Lai:2009cn}.
Moreover, at realistic
baryon densities of compact stars, the residual interaction between
strangeons could be stronger than their kinetic energy, so strangeons
would be trapped in the potential well and the bulk of the dense
matter in the compact stars are crystallized into a solid state at low temperature~\citep{Xu:2003xe, Xu:2008nd}.
SSs can account for many observational facts in
astrophysics, such as pulsar glitches, sub-pulse driftings, extremely strong
magnetic fields, the transient bursts of GCRT J1745$-$3009
\citep[see e.g. ][for details]{Xu:1999bw, Zhou:2004ue, Xu:2004sf, Yue:2006it, Zhu:2005yh, 2023AdPhX...837433L}, even  for fast radio bursts related to Galactic magnetars~\citep{2022SCPMA..6589511W,2022ApJ...927..105W}}. To characterize the strong-repulsive interaction at short distances and the
nonrelativistic nature of strangeons, a phenomenological Lennard-Jones model
with two parameters has been adopted to describe the EOS of
SSs \citep{Lai:2009cn}. Besides, the tidal deformability of merging binary SSs,
as well as the ejecta and light curves, have been discussed by
\citet{Lai:2017mjv, Lai:2018ugk, Lai:2020mlu}. Recently, \citet{Gao:2021uus}
have discussed the universal relations between the moments of inertia, the tidal
deformabilities, the quadrupole moments, and the shape eccentricity
\citep{Gao:2023mwu} of SSs.

If SSs indeed exist, they can release enough gravitational energy during
starquakes to allow successful GFs to happen \citep{Xu:2006mp, Horvath:2006ua,
Xu:2006qh}. In other words, we can adopt the asteroseismological methods to
probe the internal structure of compact stars. Some types of oscillation modes
strongly couple to the space-time continuum, and can damp on relatively short
time scales by emitting gravitational waves (GWs). \citet{Li:2022qql} have
recently studied the oscillation modes and the related GWs of SSs. They
discussed the universal relations between the fundamental ($f$)-mode frequencies
and the global properties of SSs, such as compactness and tidal deformability. 
Moreover, inverted hybrid stars are discussed in \citet{Zhang:2023zth}, and
extensions to pseudo-Newtonian gravity can be found in \citet{Li:2023ijg}.

In the present work, for the first time we study the torsional oscillation modes
of SSs in detail, and also discuss the effects of the magnetic field on the
frequencies of the torsional modes. In the SS scenarios, we attempt to explain
the observational QPO frequencies during the GFs of SGR 1806$-$20, SGR 1900+14
and SGR J1550$-$5418.  Our results suggest that the SS-model can explain the
high-frequency ($\gtrsim 150$\,Hz) QPOs, while it meets challenges of some
low-frequency QPOs,  such as the $18$\,Hz, $30$\,Hz and $92$\,Hz frequencies for
SGR 1806$-$20, and the $40$\,Hz frequency for SGR J1935+2154. We attribute these
difficulties to the large shear modulus of SSs, which could reach $10^{32}\,\rm
erg \,cm^{-3}$ \citep{Xu:2003xe}. In view of this, we further consider the
interface modes of SSs in the interface between the ocean and crust to explain
these low-frequency QPOs. Such an ocean layer could have a width in the range of
$\sim 10$--$50$\,m, consisting of a plasma of electrons and nuclei
\citep{Medin:2010kt}. The Coulomb interaction energy between ions is greater
than the thermal energy, leading to  liquid behaviors. The ocean can influence
the transport and release of thermal energy from the surface of SSs. It is in
this ocean layer that the burning that produces X-ray bursts takes place.
\citet{McDermott:1988} have investigated  non-radial oscillations of NSs with a
fluid core, solid crust, and thin surface fluid ocean. Based on such a
three-component model, they proposed a new oscillation mode called the interface
mode. This interface mode can be caused by the interface not only between the
ocean and the crust but also between the crust and the core. \citet{Piro:2004bh}
have discussed the ocean-crust interface wave, exploring its properties both
analytically and numerically for a two-component NS envelope model. We follow
these ideas and apply them to SSs. We find that the ocean-crust interface mode
of SSs can explain well the observed low-frequency QPOs in the GFs.

Furthermore, we also calculate the frequencies of the torsional modes of the
quark stars (QSs), which were firstly proposed by \citet{Witten:1984rs}.
Typically, QSs could not have torsional shear modes due to its ultra-dense quark
liquid extending up to the surface \citep{Haensel:1986}. However, QSs can have a
thin crust that extends to the neutron drip density \citep{Alcock:1986hz}.
\citet{Jaikumar:2005ne} suggested that such a crust could be made up of nuggets
of strange quark matter embedded in a uniform electron background. Although
\citet{Watts:2006hk} have calculated the torsional oscillations of QSs and
discussed the effects of the magnetic field and temperature on the frequencies
of torsional modes, their results should be modified using a more complete model
with general-relativistic corrections and magnetic fields. We calculate the
torsional oscillations for both the thin crust model and the quark nugget crust
model in this work, and discuss the effects of the magnetic field on the
frequencies. 

The paper is organized as follows. In Sec.~\ref{sec: Equilibrium Configuration},
we present our equilibrium models for  SSs. For non-magnetized and magnetized
stars, we discuss in Sec.~\ref{sec: perturbation equations} the numerical setups
for solving the perturbation equations of the torsional oscillations with the
Cowling approximation. In Sec.~\ref{sec: Numerical results}, we present the
frequencies for SSs and QSs, as well as the fitting formulae for the effects of
the magnetic field in the oscillation spectrum. Finally, we conclude in
Sec.~\ref{sec: Conclusions}. Throughout this paper, we adopt geometric units
with $c=G=1$, where $c$ and $G$ are the speed of light and the gravitational
constant, respectively.

\section{Equilibrium Configuration}\label{sec: Equilibrium Configuration}

The general-relativistic equilibrium stellar model is assumed to be spherically
symmetric and  static, as described by the TOV equations. The line element of
spacetime reads
\begin{equation}
  {\rm d} s^2 = -e^{2\Phi} {\rm d}t^2 + e^{2\Lambda} {\rm d} r^2 + r^2({\rm d}
  \theta^2 + \sin^2\theta {\rm d}\phi^2) \,,
\end{equation}
where $\Phi$ and $\Lambda$ are  functions of $r$.  Typically, for magnetars, the
magnetic energy ($E_{\rm m}$) is a few orders of magnitude smaller than the
gravitational energy ($E_{\rm g}$) with a ratio,
\begin{equation}
 \frac{E_{\rm m}}{E_{\rm g}}
    \sim \frac{B^2 R^3}{ M^2 / R}
    \sim 10^{-4} \left(\frac{B}{10^{16} \,{\rm G}}\right)^2,
\end{equation}
where $R$ and $M$ are the radius and mass of a magnetar, respectively, and $B$
is the surface magnetic field strength. In view of this, deformations on the
spherical symmetry induced by the magnetic fields are usually small for
magnetars \citep{Colaiuda:2007br, Haskell:2007bh}. Note that the general
stress-energy tensor for a magnetized relativistic star is given by, 
\begin{equation}
 T^{\mu\nu} =(\rho + P)u^{\mu}u^{\nu} + Pg^{\mu\nu} +
  H^2 u^{\mu} u^{\nu} + \frac{1}{2} H^2 g^{\mu\nu}
\label{eq: stress-energy} \,,
\end{equation}
where $\rho$ is the energy density, $P$ is the pressure, $u^{\mu}$ is the
4-velocity of  fluid, and $H^{\mu} = B^{\mu}/\sqrt{4 \pi}$ is the magnetic
field.

Based on the above setting, \citet{Sotani:2006at} investigated torsional
oscillations of relativistic stars with dipole magnetic fields.  They showed
that the magnetic field could exhibit a poloidal geometry, and can be derived by
solving the Grad-Shafranov equation,
\begin{equation}
 \frac{{\rm d^2} a_{1}}{{\rm d} r^2}+ \left(\Phi' - \Lambda' \right)
\frac{{\rm d}  a_{1}}{{\rm d} r}- \frac{2}{r^2}e^{2\Lambda}a_{1} 
=-4\pi e^{2\Lambda} c_{0}r^2(\rho+P) \label{eq: MF} \,,
\end{equation}
where $a_{1}$ is the radial component of the electromagnetic four-potential, and
$c_{0}$ is a constant.  We use primes to represent the radial derivatives
hereafter.  To solve Eq.~(\ref{eq: MF}), we must specify the boundary conditions
at both the center and the surface of a star.  Regularity of the solution at the
origin requires $a_{1}=\alpha_{0}r^2$, where $\alpha_{0}$ is a constant.  At the
surface, the internal solutions must be consistent with those above the surface
with an external magnetic field.  We consider a dipole field in the vacuum in
this work.  Therefore, the solution above the surface is given by,
\begin{equation}
 a_1 = -\frac{3\beta_{0}}{8M^3}r^2 \left[\ln\left(1-\frac{2M}{r}\right)
 + \frac{2M}{r} + \frac{2M^2}{r^2}\right] \,,
\end{equation}
where $\beta_{0}$ is the magnetic dipole moment.  The corresponding magnetic
field thus has the form \citep{Konno:1999zv},
\begin{align}
B_{r} &= \frac{2 \cos\theta }{r^2}a_{1} \,, \\
B_{\theta} &= -\frac{e^{-\Lambda} \sin\theta }{r}\frac{{\rm d}  a_{1}}{{\rm d} r} \,.
\end{align}

\begin{figure}
    \centering 
    \includegraphics[width=8cm]{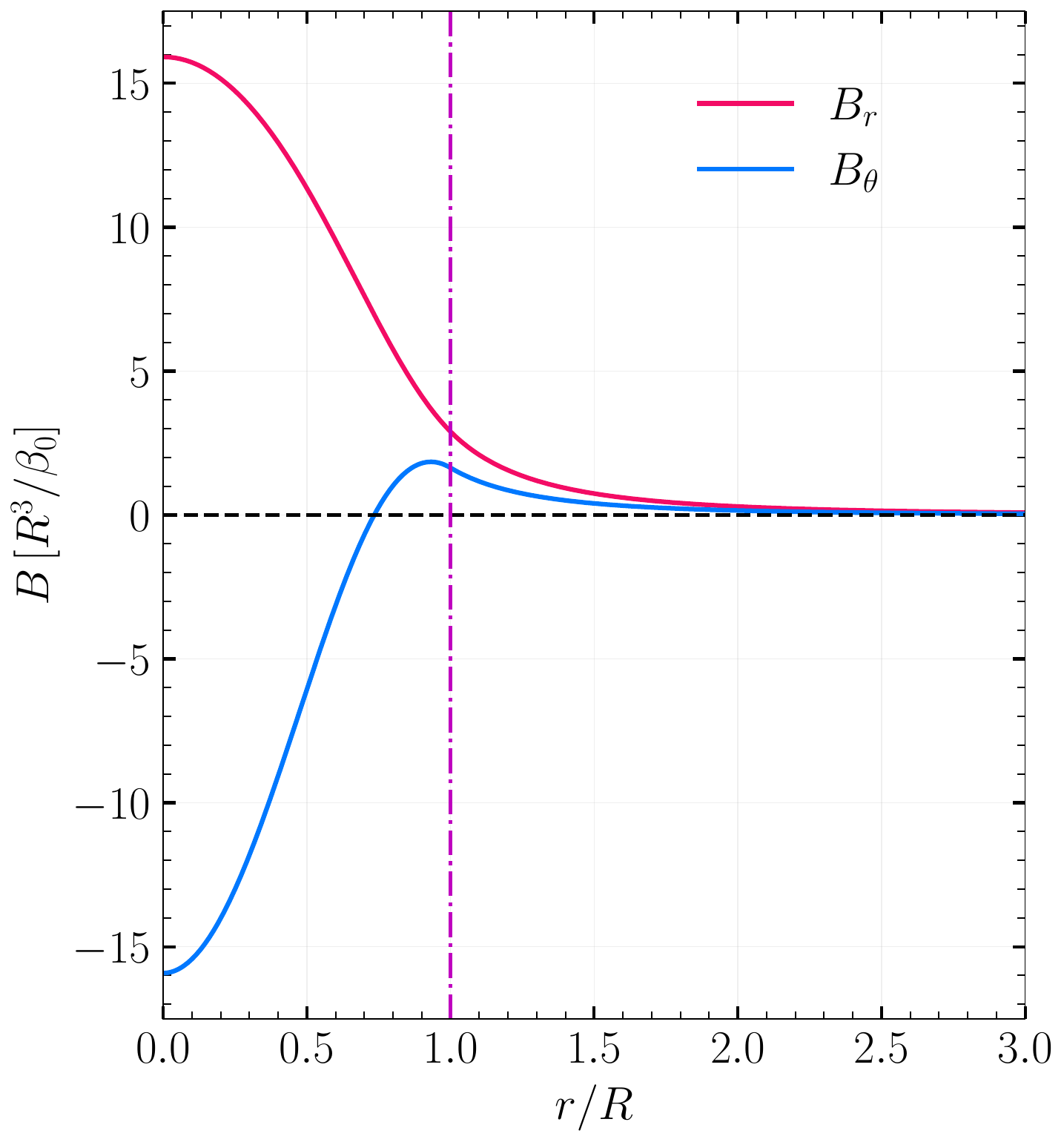}
    \caption{Profiles of the magnetic field components against the radial
    coordinate $r$. $B_{r}$ and $B_{\theta}$ are evaluated at $\theta =0$ and
    $\theta =\pi/2$, respectively. We set the compactness $\mathcal C=0.2$. The
    purple dotted-dashed line denotes the surface of the star.} 
    \label{fig: B_value}
\end{figure}

We plot in Fig.~\ref{fig: B_value} the profiles of the magnetic field
components, $B_{r}$ and $B_{\theta}$, against the radial coordinate $r$.  The
magnetic fields are normalized by $\beta_{0}/R^3$.  In this plot, we have used
the polytropic EOS, $P=k\rho^{\gamma}$, with $\gamma=2$ and a compactness 
$\mathcal C=0.2$ for the $k$ value. 

\section{Torsional oscillations in the Cowling approximation}
\label{sec: perturbation equations}

Axial and polar perturbations do not couple with each other when we consider the
pure axisymmetric perturbations on a spherically symmetric star.  Here we
consider only the torsional oscillations of a non-magnetized relativistic star
with axial perturbations.  The density variations in the spherically symmetric
star would not be induced.  For this reason, we neglect the perturbations of
spacetime using the Cowling approximation.  The axial perturbation equations for
the elastic solid star in the Cowling approximation is written as
\citep{Samuelsson:2006tt, Sotani:2012qc},
\begin{multline}
	Y''  +  \left( \frac{4}{r}+\Phi'-\Lambda'+\frac{\mu'}{\mu} \right) Y' \\
 + \left[ \frac{\rho+P}{\mu}\,\omega^2 {\rm e}^{-2\Phi}
  -\frac{(\ell+2)(\ell-1)}{r^2}\right]{\rm e}^{2\Lambda}Y = 0 \,,
   \label{eq: fc_1}
\end{multline}
where $\mu$ is the shear modulus, $\omega$ is the angular frequency, $Y(r)$
describes the radial part of the angular oscillation amplitude, and the integer
$\ell$ is the angular separation constant which enters when $Y(r)$ is expanded
in spherical harmonics $Y_{\ell m} (\theta, \phi)$. 

Now we extend our studies to the torsional oscillation of a magnetized
relativistic star. \citet{Sotani:2006at} derived the perturbation equations of
the magnetized relativistic star using the relativistic Cowling approximation.
The final perturbation equation is,
\begin{equation}
A_{\ell}(r)Y'' +B_{\ell}(r)Y'+C_{\ell}(r)Y=0 \,,
\label{eq: fc_2}
\end{equation}
where the coefficients are given in terms of the functions describing the
equilibrium metric, fluid, and the magnetic field of the star,
\begin{align}
 A_{\ell}(r) &= \mu + (1 + 2 \lambda_1)\frac{{a_1}^2}{\pi r^4} \,, 
 \label{eq: coefficients_1} \\
B_{\ell}(r) &=\left(\frac{4}{r} + \Phi' - \Lambda'\right)\mu + \mu'  \nonumber\\ 
      & \quad+ (1 + 2\lambda_1)\frac{a_1}{\pi r^4}\left[\left(\Phi' - \Lambda'\right)a_1
     + 2{a_1}'\right] \,,
\label{eq: coefficients_2} \\
C_{\ell}(r) &=\left[\left(\rho + P + (1 +2\lambda_1)\frac{{a_1}^2}{\pi r^4}
\right)e^{2\Lambda}
     - \frac{\lambda_1 {{a_1}'}^2}{2\pi r^2}\right]\omega^2 e^{-2\Phi}  \nonumber\\
    & \quad - (\lambda-2)\left(\frac{ \mu e^{2\Lambda}}{r^2}
- \frac{\lambda_1{{a_1}'}^2}{2\pi r^4}\right) \nonumber\\
     & \quad + (2 + 5\lambda_1)\frac{a_1}{2\pi r^4}\left\{\left(\Phi'
- \Lambda'\right){a_1}' + {a_1}''\right\} \,,
\label{eq: coefficients_3} 
\end{align}
where $ \lambda = \ell(\ell+1)$, and $\lambda_1 = -
\ell(\ell+1)/[(2\ell-1)(2\ell+3)]$.  To solve Eqs.~(\ref{eq: fc_1}) and
(\ref{eq: fc_2}) and determine the oscillation frequencies, the boundary
conditions require that the traction vanishes at the top and the bottom of the
crust. In the next section, we will use perturbation equations (\ref{eq: fc_1})
and (\ref{eq: fc_2}) to study torsional oscillation modes of the SSs and discuss
possible effects from the magnetic field.

\section{NUMERICAL RESULTS}\label{sec: Numerical results}

We present results of torsional oscillation modes for SSs in Sec.~\ref{sec:
SSs}, and make comparisons to QSs in Sec.~\ref{sec: QSs}.

\begin{table*}
    \centering
    \caption{Frequencies (in the unit of Hz) of the fundamental torsional modes
    ${}_{\ell} f_{0}$ for SSs without the magnetic fields (i.e. $n=0$ and
    $B=0$). The subscript in the models denote the SS mass; taking SS$_{12}$ as
    an example, the subscript ``12'' represents $M = 1.2\,M_{\odot}$. Values in
    bold are close to the QPO observations (within $2\%$) for SGR 1806$-$20, SGR
    1900+14 and SGR J1550$-$5418.}
    \renewcommand\arraystretch{1.5}
    \begin{tabular}{c c c c c c c c c c}
    \hline
    Model &  $\ell=2$ & $\ell=3$ & $\ell=4$ & $\ell=5$ & $\ell=6$ 
    & $\ell=7$ & $\ell=8$ & $\ell=9$ & $\ell=10$ \\
    \hline
    SS$_{12}$ & 277    & 431.3 & 572.1 & 707.2  & 839.3 & 969.4 & 1098.3 & 1226.3 & 1353.6 \\
    SS$_{14}$ & \bf253.3 & 394.9 & 524.2 & 648.6  & 770.2 & 890    & 1008.9 & 1126.9 & 1244.3 \\
    SS$_{16}$ & 233    & 363.7 & 483.3 & 598.3  & 710.9 & 822.1 & 932.3   & 1041.8 & 1150.8 \\
    SS$_{18}$ & 214.8 & 335.7 & 446.5 & 553.1  & 657.9 & 761.2 & 863.6   & 965.5   & 1067 \\
    SS$_{20}$ & 198.4 & 310.4 & 413.3 & 512.5  & 609.9 & 706.1 & 801.5   & 896.4   & 991 \\
    SS$_{22}$ & 183.5 & 287.5 & 383.2 & 475.6  & 566.3 & 656    & 745.1   & 833.7   & 922 \\
    SS$_{24}$ & 169.7 & \bf266.3 & 355.4 & 441.5  & 526    & 609.7 & 692.8   & 775.5   & 858 \\
    SS$_{26}$ & \bf155.6 & 244.6 & 326.7 & 406.2  & 484.4 & 561.7 & \bf638.6   & 715.1   & 791.4 \\
    SS$_{28}$ & \bf145    & 228.1 & 305    & 379.5  & 452.7 & 525.2 & 597.2       & 668.9   & 740.3 \\
    \hline
    \end{tabular}
    \label{tab: Table_1}
\end{table*}

\begin{table*}
    \centering
    \caption{Same as Table~\ref{tab: Table_1}, but for  the first overtone
    $(n=1)$ of the torsional modes ${}_{\ell} f_{1}$ of SSs.}
    \renewcommand\arraystretch{1.5}
    \begin{tabular}{c c c c c c c c c c}
    \hline
    Model &  $\ell=2$ & $\ell=3$ & $\ell=4$ & $\ell=5$ & $\ell=6$ 
    & $\ell=7$ & $\ell=8$ & $\ell=9$ & $\ell=10$ \\
    \hline
    SS$_{12}$ & 683.3 & 820.8 & 956.9 & 1090.6 & 1223    & 1354.3  & 1484.5 & 1614.3 & 1744.2  \\
    SS$_{14}$ & \bf612.3 & 737.7 & 861    & 982.7   & 1103.4 & 1223.4  & 1342.7 & 1457.9 & 1579.8  \\
    SS$_{16}$ & 551.5 & 666    & 778.8 & 890.4   & 1001.1 & 1111.2  & 1220.9 & 1330.1 & 1439.1  \\
    SS$_{18}$ & 497.3 & 602    & 705.3 & 807.6   & 909.6   & 1010.9 & 1101.1  & 1212.6 & 1313.1  \\
    SS$_{20}$ & 448.3 & 544.3 & \bf639.1 & 733.2   & 826.9   & 920.3   & 1013.4  & 1106.4 & 1193.3  \\
    SS$_{22}$ & 404.3 & 492.2 & 579.3 & 666      & 752.4   & 838.6   & 924.6    & 1010.6 & 1096.5 \\
    SS$_{24}$ & 363.9 & 444.5 & 524.5 & 604.2   & 683.8   & 763.4   & 842.9    & 922.5   & 1002 \\
    SS$_{26}$ & 322.8 & 395.8 & 468.5 & 541.2   & \bf613.9   & 686.6   & 759.4    & 832.4   & 905.4 \\
    SS$_{28}$ & 292.1 & 359.4 & 426.7 & 494      & 561.5       & 629      & 697       & 765      & 833.1 \\
    \hline
    \end{tabular}
    \label{tab: Table_2}
\end{table*}

\begin{table*}
    \centering
    \caption{Same as Table~\ref{tab: Table_1}, but for the higher overtones
    ${}_{3} f_{n}$ of the torsional modes of SSs for $\ell=3$.}
    \renewcommand\arraystretch{1.5}
    \begin{tabular}{c c c c c c c}
    \hline
    Model &  $n=2$ & $n=3$ & $n=4$ & $n=5$ & $n=6$ & $n=7$ \\
        \hline
        SS$_{12}$ & 820.8 & 1136.2 & 1434.4  & 1742.6 & 2042.2      & 2340.5 \\
        SS$_{14}$ & 737.7 & 1016.6 & 1283.5  & 1558.2 & \bf1821.2 & 2091.5 \\
        SS$_{16}$ & 665.9 & 916.3   & 1159.7  & 1400.7 & 1640.2      & \bf1878.7 \\
        SS$_{18}$ & 602    & 825.7   & 1044     & 1260.2 & 1474.8      & 1689.1 \\
        \hline
    \end{tabular}
    \label{tab: Table_3}
\end{table*}

\subsection{Strangeon stars}\label{sec: SSs}

\citet{Xu:2003xe} conjectured that cold quark matter with very high baryon
density could be in a solid state, and considered a SS at low temperature should
be a solid star. The shear modulus of solid quark matter could be $\sim
10^{32}\,\rm erg \,cm^{-3}$\citep{Xu:2003xe}, much larger than those of typical
NSs \citep[$\sim 10^{30}\,\rm erg \,cm^{-3}$; ][]{Duncan:1998my, Piro:2005jf}.

We first exhibit in Table~\ref{tab: Table_1} the frequencies of the SS
fundamental modes ${}_{\ell} f_{0}$ for $\ell=2$ to $10$. It shows that the
frequency of $\ell=2$ mode varies from $145$ to $277$\,Hz, depending on the SS
mass. Given the SS mass, the frequencies tend to increase with a higher $\ell$.
In comparison, for NSs, the frequencies of the fundamental $\ell=2$ mode only
range from $17$ to $29$\,Hz \citep{Sotani:2006at}.  Therefore, the frequency
range of the $\ell=2$ mode of SSs is much larger than that of NSs. This is
introduced by the larger shear modulus of SSs.

We show in Table~\ref{tab: Table_2} the frequencies of the first overtone
${}_{\ell} f_{1}$ for SSs. We find that the frequencies  range from $300$\,Hz to
$1700$\,Hz. Table~\ref{tab: Table_2} shows that the frequency of the first
overtone decreases as the SS mass increases, which is the same as the
fundamental modes in Table~\ref{tab: Table_1}. However, such an anti-correlation
differs from those of NSs, where the frequency of the first overtone would
increase with a higher NS mass \citep{Sotani:2006at}. Note that such
anti-correlation can also be found in higher overtones (see Table~\ref{tab:
Table_3}).

\Reply{
In order to compare with the torsional mode frequency of NSs, 
we calculate the torsional mode frequency of SSs using the same mass range of $1.2 M_{\odot}$--$2.8 M_{\odot}$. However, the $2.8 M_{\odot}$ value is not the SSs' maximum mass. 
As described in the Introduction, in a reasonable model, the EOS of SSs is
completely determined by the depth of the potential and the number
density of baryons at the surface of the star. 
Because the strangeons are nonrelativistic there is a very strong repulsion
at a short inter-cluster distance~\citep{Lai:2009cn, Lai:2017ney, Gao:2021uus, Li:2022qql}, which could lead to the maximal mass of SSs over $3\,M_{\odot}$.
}

In Tables~\ref{tab: Table_1}, \ref{tab: Table_2} and \ref{tab: Table_3}, we have
marked the frequencies that are good fit (within $2\%$) to the observed QPO
frequencies for some SGRs.  For examples, we have: model SS$_{26}$ (whose
${}_{2} f_{0}=155.6$\,Hz) for SGR 1900+14, model SS$_{24}$ (whose ${}_{3}
f_{0}=266.3$\,Hz) for SGR J1550$-$5418, and model SS$_{14}$ (whose ${}_{2}
f_{1}=612.3$\,Hz) for SGR 1806$-$20.  Based on our results, the observed high
frequencies of $150$\,Hz, $625$\,Hz, and $1837$\,Hz might correspond to ${}_{2}
f_{0}$, ${}_{2} f_{1}$, and ${}_{3} f_{6}$, respectively.  We also find that the
SS model could explain well the high-frequency QPOs in the GFs of SGR 1806$-$20,
SGR 1900+14 and SGR J1550$-$5418. 

However, there appears to have difficulty when using the SS model to interpret
the low-frequency QPOs (e.g., $18$\,Hz and $29$\,Hz for SGR 1806$-$20).  For
this reason, we propose that SSs may have a thin surface ocean with density and
temperature in the range of $10^{6}$--$10^{9}\,\rm g \, cm ^{-3}$ and
$10^{8}$--$10^{9}$\,K, respectively. With such an ocean layer, we can use the
interface modes of the ocean-crust interface in order to explain the
low-frequency QPOs. The frequency of the interface mode can be analytically
approximated by \citep{Piro:2004bh},
\begin{align}
       f \approx & \, 16.5\textrm{\,Hz}
		\left(\frac{\Gamma}{173}\right)^{1/2}
		\left(\frac{T_8}{4}\right)^{1/2}
		\nonumber
		\\
		& \, \times\left(\frac{64}{A}\right)^{1/2}
		\left(\frac{10\textrm{ km}}{R}\right)
		\left[\frac{\ell(\ell+1)}{2}\right]^{1/2}\,,
		\label{eq: f_1}
\end{align}
where $T_8\equiv T/10^8\textrm{K}$, $A$ is the baryon number, and $\Gamma$ is a
dimensionless parameter that determines the liquid-solid transition via,
\begin{align}
	\Gamma \equiv \frac{(Ze)^2}{ak_{\rm B}T}
	= \frac{127}{T_8/4}\left(\frac{Z}{30}\right)^2
	\left( \frac{64}{A}\right)^{1/3}
	\left(\frac{\rho}{10^9\textrm{ g cm}^{-3}} \right)^{1/3}\,,
\end{align}
where $k_{\rm B}$ is the Boltzmann's constant, and $Z$ is the proton number. The
crystallization point occurs at $\Gamma = 173$ \citep{Farouki:1993}. Using
Eq.~(\ref{eq: f_1}), we calculate the frequencies of the interface modes with
different $\ell$. The SS mass and radius are fixed at $M = 1.4\, M_{\odot}$ and
$R = 10$\,km, respectively. According to Eq.~(\ref{eq: f_1}), we see that the
frequency is independent of the number of the overtone. It is not difficult to
verify that the ocean-crust interface modes could interpret the recorded
low-frequency QPOs, such as ${}_{1} f=16.3$\,Hz, ${}_{2} f=28.5$\,Hz, and
${}_{8} f=99$\,Hz for SGR 1806$-$20, ${}_{4} f=52.1$\,Hz and ${}_{7} f=87.3$\,Hz
for SGR 1900+14, and ${}_{3} f=40.4$\,Hz for SGR J1935+2154.

As already emphasized by \citet{Sotani:2006at}, the shift in the frequencies
would be significant when the magnetic field exceeds $\sim 10 ^{15}$\,G. 
Following \citet{Sotani:2006at}, now we discuss the effects of the magnetic
fields. In the presence of magnetic fields, frequencies are shifted as,
\begin{equation}
{}_\ell f_n = {}_\ell f_n^{(0)} \left[ 1+ {}_\ell \alpha_n
\left(\frac{B}{B_{\mu}}\right)^2 \right]^{1/2}\,,
\label{eq: fit}
\end{equation}
where ${}_\ell {\alpha}_n$ is a coefficient depending on the structure of the
star, and ${}_\ell f_n^{(0)}$ is the frequency of the non-magnetized star. Note
that the typical magnetic field strength is defined as $B_\mu \equiv (4 \pi
\mu)^{1/2}$. More details can be found in \citet{Messios:2001br}. 

\begin{figure}
    \centering 
    \includegraphics[width=8cm]{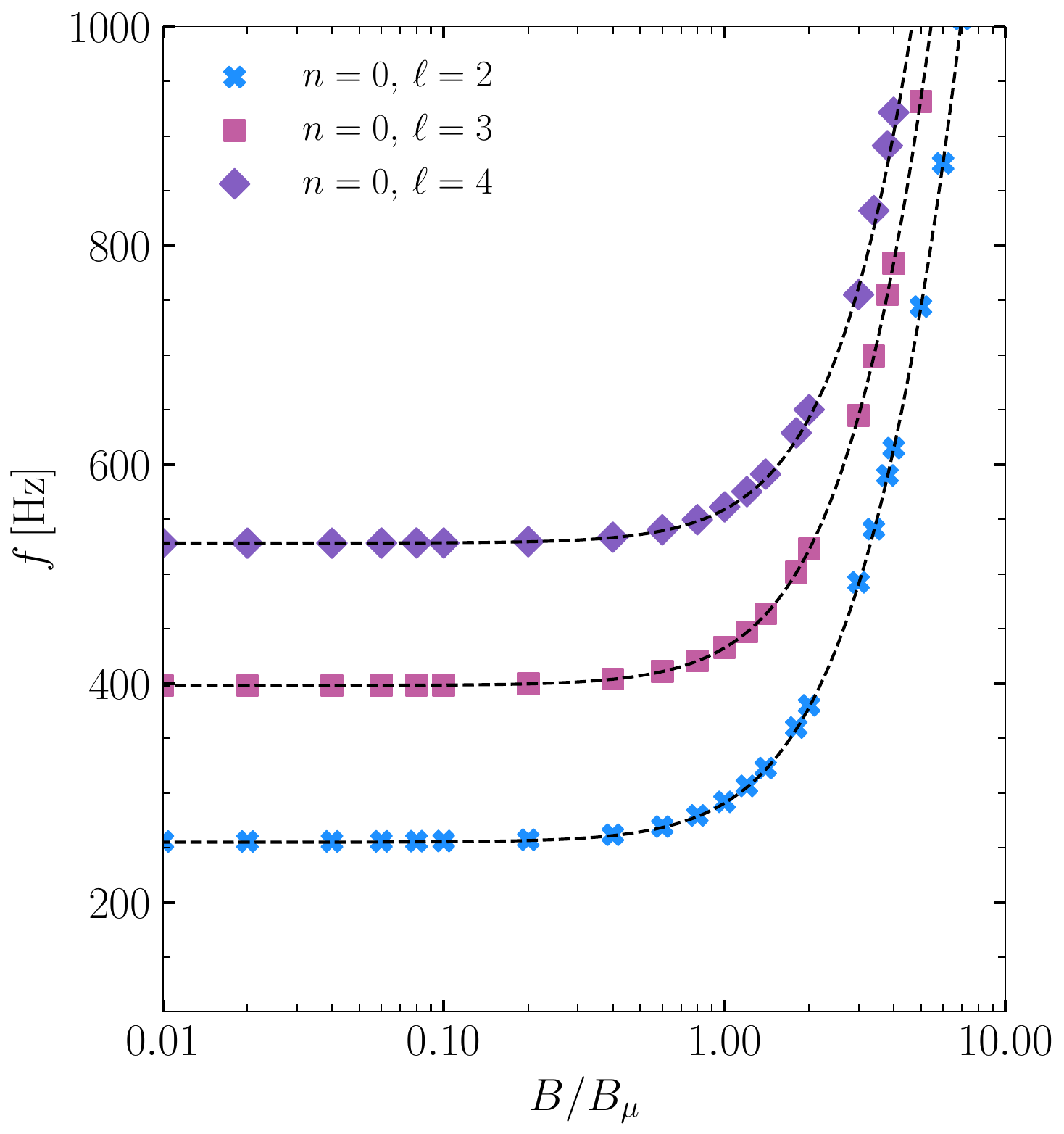}
    \caption{Frequencies of the fundamental $n=0$ modes with $\ell=2$, $\ell=3$,
    and $\ell=4$ as a function of the magnetic field. We set the SS mass to be
    $M = 1.4\, M_{\odot}$. Individual numerical results are denoted with
    different marks in different colours. The dashed lines correspond to the
    empirical formula~(\ref{eq: fit}) with different coefficient values. The
    fitting coefficients ${}_\ell {\alpha}_n$ are $0.3$, $0.18$, and $0.12$ for
    $\ell=2$, $\ell=3$, and $\ell=4$, respectively.}
    \label{fig: f_SS_1} 
\end{figure}

\begin{figure}
    \centering 
    \includegraphics[width=8cm]{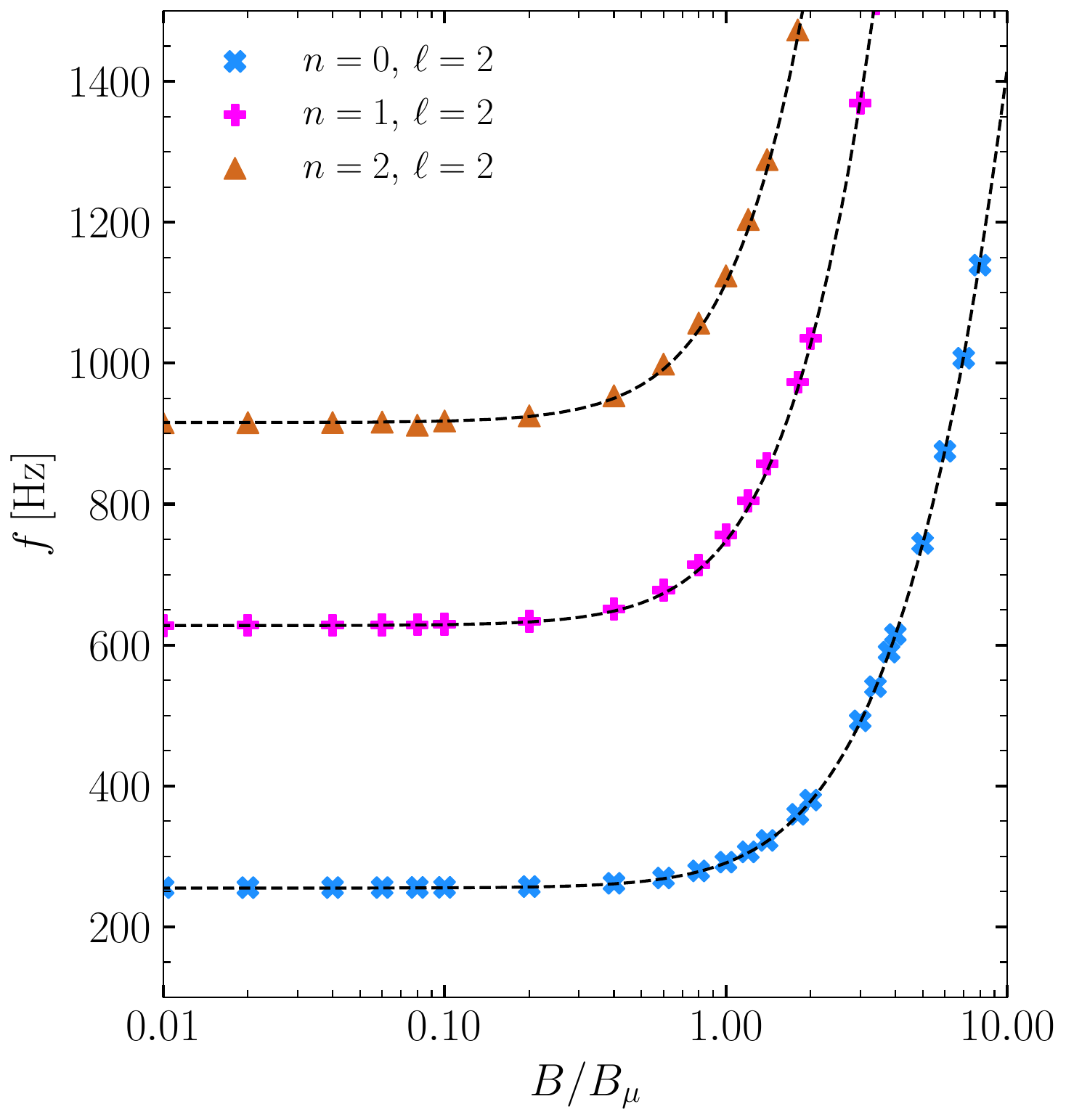}
    \caption{Frequencies of different overtones as a function of the normalized
    magnetic field for $\ell=2$. The dashed lines are our fits using
    Eq.~(\ref{eq: fit}). The SS mass is $M = 1.4\, M_{\odot}$. The coefficients
    ${}_\ell {\alpha}_n$ are $0.3$, $0.42$ and $0.48$ for $n=0$, $n=1$, and
    $n=2$, respectively. }
    \label{fig: f_SS_2}
\end{figure}

In Figs.~\ref{fig: f_SS_1} and \ref{fig: f_SS_2}, we show the effects of the
magnetic field on the frequencies of the torsional modes. The magnetic field
strength is normalized by $B_{\mu} = 4 \times 10^{16}$\,G. Different dashed
lines in Figs.~\ref{fig: f_SS_1} and \ref{fig: f_SS_2} are our fits to the
calculated numerical data with a high accuracy. For $B > B_{\mu}$, we find that
the frequencies follow a quadratic increase against the magnetic field, and tend
to become less sensitive to the SS parameters. NSs could have similar behaviors,
but the turning-point value of the magnetic field strength is much lower
\citep[$\sim 4 \times 10^{15}$\,G;][]{Sotani:2006at}.


\subsection{Comparisons with quark stars}\label{sec: QSs}

\begin{figure}
    \centering 
    \includegraphics[width=8cm]{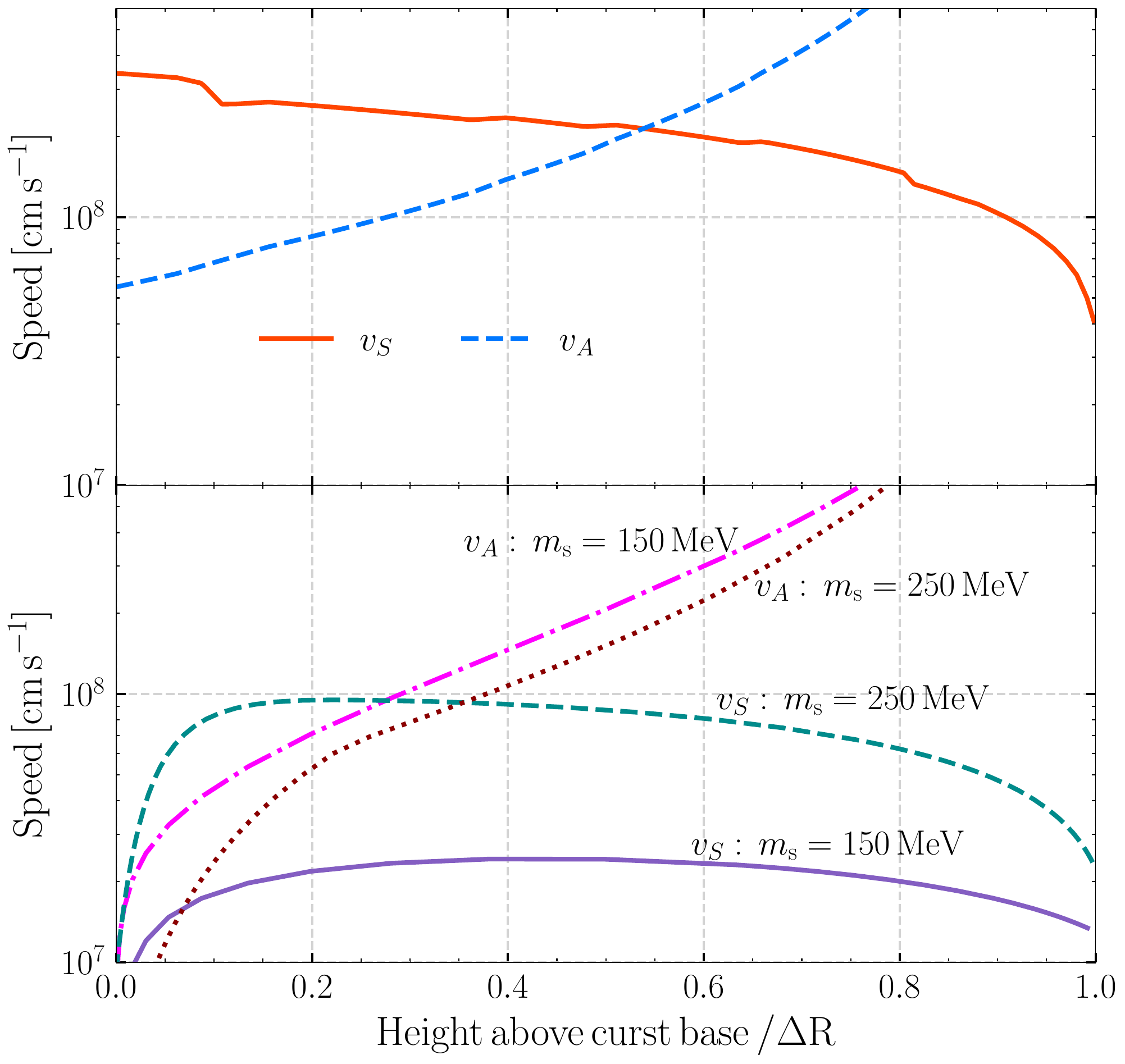}
    \caption{Shear speed $v_{\rm S}$ and  alfv\'en speed $v_{\rm A}$ in the
    crust of stars for a QS with a fixed mass  $M = 1.4\, M_{\odot}$, a radius
    $R = 12 \, \rm km$ , and a magnetic field  $B = 10^{14}\, \rm G$. The upper
    panel is for the thin nuclear crust, and the lower panel is for the crust
    with nuggets. Note that in the lower panel, we consider $m_{\rm s} = 150 \,
    \rm MeV$ and $m_{\rm s} = 250 \, \rm MeV$ denoted with different lines in
    different colors.  }
    \label{fig: speed}
\end{figure}

The frequencies of torsional oscillations depend sensitively on the property of
the crust. Considering there is no solid region for bare QSs, they could not
account for the torsional shear oscillations. However, \citet{Alcock:1986hz}
suggested that  QSs could have a thin nuclear crust that extends to the neutron
drip density (i.e. $\rho \approx 4 \times 10^{11} \rm g\, cm^{-3}$).
\citet{Jaikumar:2005ne} proposed another possible model that a crust is made up
of  nuggets of strange quark matter embedded in a uniform electron background. 
In this subsection, we present detailed calculations of both models.

Following \citet{Watts:2006hk}, we use the standard general relativistic
algorithm with a shear speed $v_{S} = (\mu/\rho)^{1/2}$ and an alfv\'en speed
$v_{A} = B/(4\pi \rho)^{1/2}$. We show in Fig.~\ref{fig: speed} the depth of the
crust for different crust models.\footnote{Note that the magnetic field is
constant ($B=10^{14}$\,G) in Fig.~\ref{fig: speed}. When we discuss the effects
of the magnetic field on the frequencies of torsional oscillations, we calculate
the torsional oscillation frequencies using Eq.~(\ref{eq: fc_2}).}  We find that
the shear speed in the nugget crust is smaller than that in the thin nuclear
crust, which is consistent with the results at a constant pressure $ v_{\rm S}
\sim \sqrt{Z^{5/3}/A}$ in \citet{Watts:2006hk}, where $A$ denotes the baryon
number. In particular, Fig.~\ref{fig: speed} shows that both $Z$ and $Z/A$ of
the nuggets decrease rapidly with the depth. In our calculation, for the thin
nuclear crust models, the EOS of QSs is described by the MIT bag model, and the
EOS of the crust is given by \citet{BPS:1971}. The shear modulus $\mu$ is,
\begin{equation}
\mu= 0.1194 \frac{n_i (Ze)^2}{a}\,,
\label{eq: mu}
\end{equation}
where $n_i$ is the ion number density, $a=[3/(4\pi n_i)]^{1/3}$ is the average
ion spacing, and $+Ze$ is the ion charge.

For the thin nuclear crust models, the frequency of the fundamental torsional
mode with $\ell=2$, ${}_{2} f_{0}$, is 64\,Hz for a given mass $M = 1.4\,
M_{\odot}$.  We find that the frequency of the fundamental torsional mode ranges
from $26$ to $64$\,Hz. To obtain a frequency $\leq 30$\,Hz requires a high mass
($M \geq 2.4\, M_{\odot}$ for all radii).  The overtone frequencies are even
higher for thin nuclear crust models.  When $B>B_{\mu}=4 \times 10^{13}\,\rm G$
the modes change the character and become dominated by the magnetic field. 

\begin{figure}
    \centering 
    \includegraphics[width=8cm]{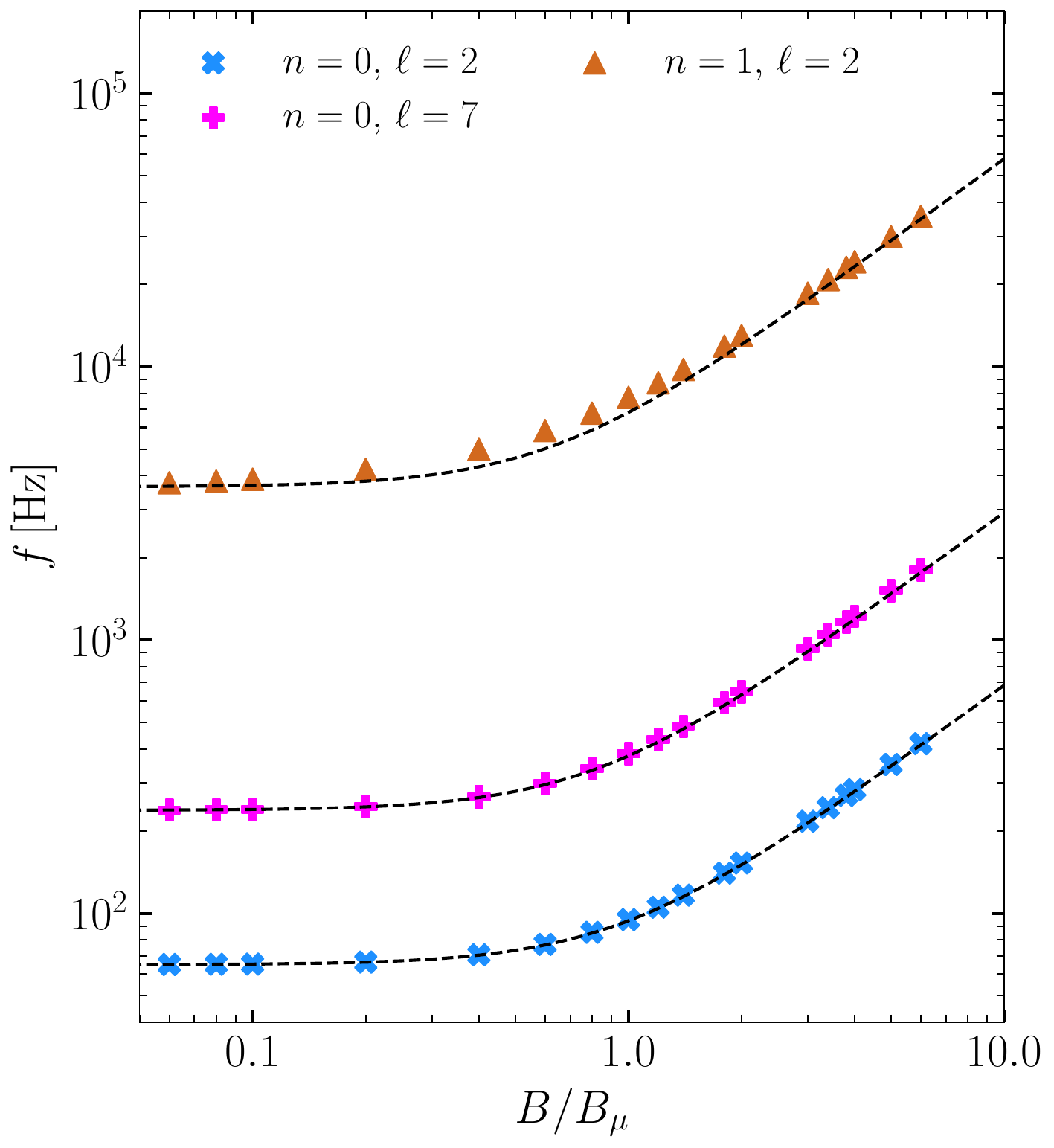}
    \caption{\Reply{Torsional mode frequencies in the QS thin nuclear crust model as a
    function of the magnetic field. The dashed lines correspond to our fits
    using the empirical formula (\ref{eq: fit}) with different coefficient
    values. The fitting values are $1.1$, $1.5$, and $2.5$ for ${}_{2} f_{0}$,
    ${}_{7} f_{0}$, and ${}_{2} f_{1}$, respectively.}}
    \label{fig: f_nuclear}
\end{figure}

In Fig.~\ref{fig: f_nuclear}, we show the effects of the magnetic field on the
torsional mode frequencies. The frequencies of the fundamental modes have
increased by up to $35\%$, which is similar to NSs.  The frequencies of the
first overtone have increased by up to $24\%$.  The typical value for magnetic
field is $B_\mu \equiv (4 \pi \mu)^{1/2}$, which depends on the crust EOS and
the shear modulus.  In particular, for the crust EOS that was described by
\citet{BPS:1971}, the shear modulus could be $\sim 10^{28}\, \rm erg\, cm^{-3}$.

\begin{figure}
    \centering 
    \includegraphics[width=8cm]{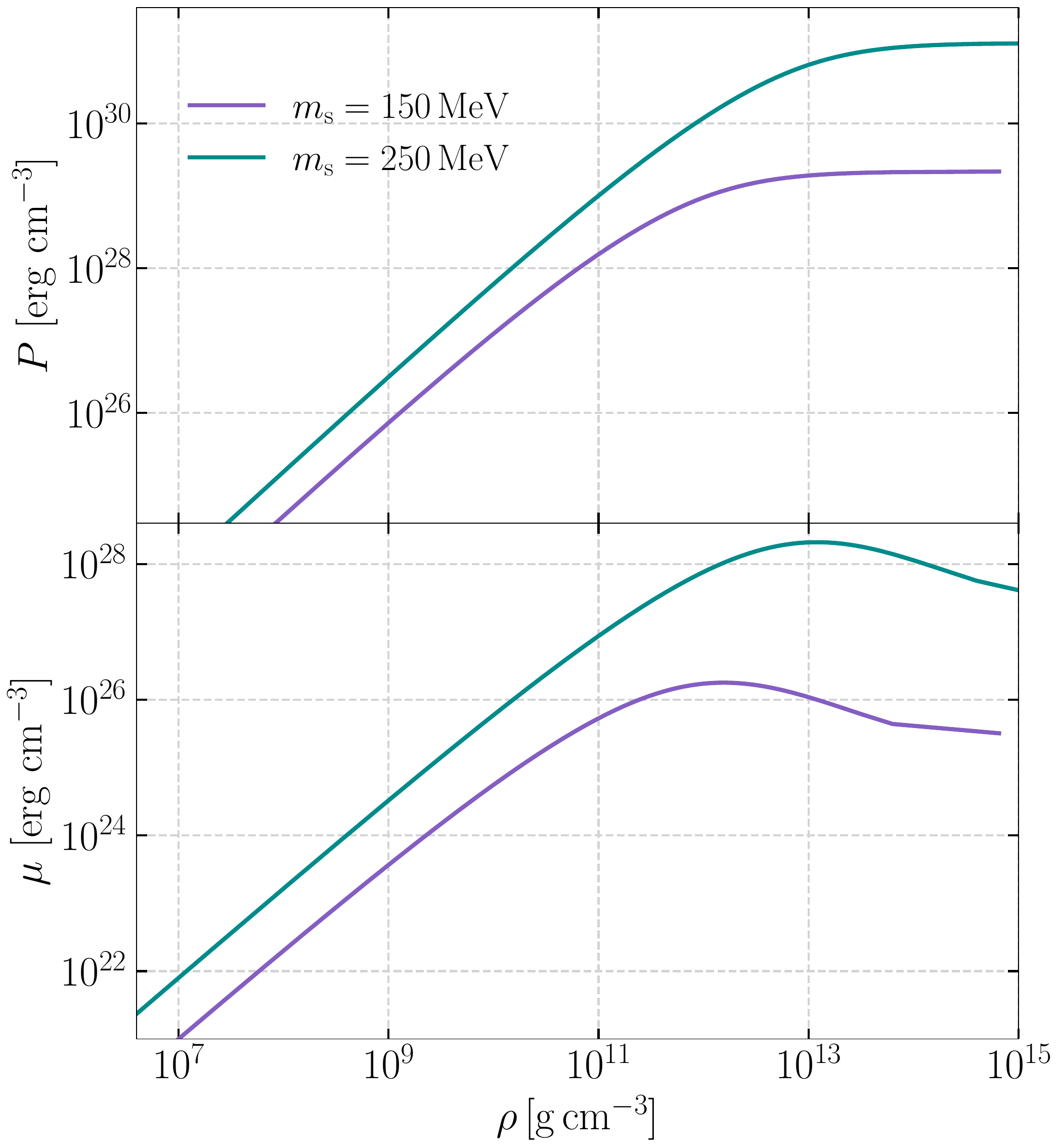}
    \caption{The upper panel shows the EOS of QSs in the nugget crust model. The
    lower panel shows the relationship between the shear modulus and the energy
    density.}
    \label{fig: EOS_nugget}
\end{figure}

For the nugget crust models, we show in Fig.~\ref{fig: EOS_nugget} the relations
of the pressure and the shear modulus against energy density. Figure~\ref{fig:
EOS_nugget} shows that the EOS and shear modulus are sensitive to the strange
quark mass, $m_{\rm s}$. The number $A$ and $Z/A$ of nuggets decrease rapidly as
the energy density increases. Using typical quark model parameters, i.e., the
MIT bag constant $B = 65 \,\rm MeV\, fm^{-3}$ and  $m_{\rm s} =150 \, \rm MeV$,
we find that QSs can have a crust width of $\Delta R= 40 $\,m for a given mass
$M = 1.4\, M_{\odot}$ with a radius $R = 10$\,km. We find that the frequency of
the fundamental mode and the first overtone is $5.11$\,Hz and $4282$\,Hz,
respectively. Additionally, for $m_{\rm s} = 250\, \rm MeV$, we find that the
thickness $\Delta R$ is $217$ m, and the frequency of the fundamental mode and
the first overtone is $7. 49$\,Hz and $2678$\,Hz, respectively. These results
could explain some QPOs in the range of $18$--$150$\,Hz but it appears to be
difficult to explain the high frequencies of $625$\,Hz and $1837$\,Hz.

\begin{figure*}
    \centering 
    \includegraphics[width=12cm]{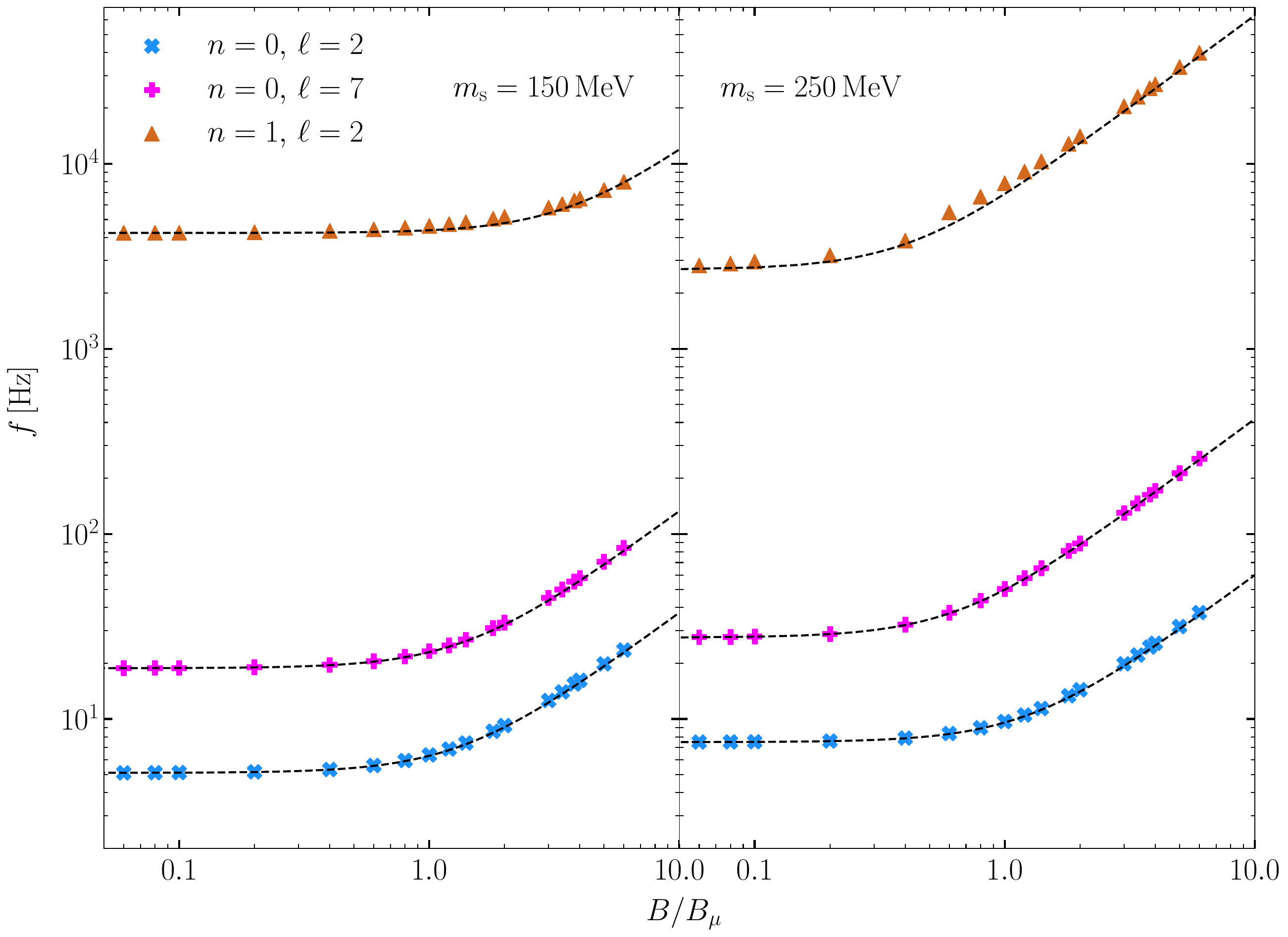}
    \caption{\Reply{Frequencies of the fundamental $n=0$ modes with $\ell = 2 \,, 7$
    and the first overtone in the nugget crust model. The left panel shows the
    results with strange quark mass $m_{\rm s}= 150 $\,MeV and the normalized
    magnetic field strength $B_{\mu} = 4 \times 10^{11}$\,G, while  for the
    right panel $m_{\rm s}= 250 $\,MeV and $B_{\mu} = 4 \times 10^{14}$\,G. The
    lines correspond to our fits using the empirical formula (\ref{eq: fit})
    with different coefficient values.}}
    \label{fig: f_nugget}
\end{figure*}

In the left panel of Fig.~\ref{fig: f_nugget}, we show the relations between the
torsional mode frequencies in the nugget crust model.   We adopt $m_{\rm s} =
150$\,MeV and a normalized magnetic field $B_{\mu} = 4 \times 10^{11}$\,G. For
this case when $B=B_{\mu}$ the frequencies of fundamental mode have increased by
up to $25 \%$, compared to the frequencies of non-magnetized models.  Using the
empirical formula (\ref{eq: fit}), we fit the coefficients for the effects of
the magnetic field. The fitting coefficient values are $0.52$, $0.49$, and
$0.07$ for ${}_{2} f_{0}$, ${}_{7} f_{0}$, and ${}_{2} f_{1}$, respectively. In
the right panel of Fig.~\ref{fig: f_nugget}, we show the results with $m_{\rm s}
= 250$\,MeV and $B_{\mu} = 4 \times 10^{14}$\,G. The fitting coefficient values
are $0.6$, $2.3$, and $5.6$ for ${}_{2} f_{0}$, ${}_{7} f_{0}$, and ${}_{2}
f_{1}$, respectively. In both panels of Fig.~\ref{fig: f_nugget}, we find that,
as expected, when the magnetic field strength is close to $B_{\mu} = 4 \times
10^{14}$\,G, the frequencies of the fundamental mode and the first overtone
significantly increases.

\Reply{
Using a plane-parallel geometry, \citet{Watts:2006hk} calculated the torsional
oscillation of QSs, and discussed the effects of the magnetic field and
temperature on the torsional mode frequencies.  Similarly, in our work, we adopt
the same EOS of the quark star and thin crust.  However, the magnetic field is a
constant in the work of \citet{Watts:2006hk}.  Differently, we consider a
relativistic star with dipole magnetic fields, calculate the frequencies of
torsional modes, and study the effects of magnetic fields.  Compared with
earlier results, when considering the magnetic field strength $B_{\mu}$, the
frequencies have increased by $25 \%$--$35 \%$ compared to the frequencies of
non-magnetized models \citep[see Figs.~\ref{fig: f_nuclear} and \ref{fig:
f_nugget} in this paper and the left panels of Figs. 2 and 3 in
][]{Watts:2006hk}.
}

\section{Discussions and Conclusions}\label{sec: Conclusions}

In this work we first studied the torsional oscillation modes of SSs using the
Cowing approximation with no magnetic field on the equilibrium configuration.
According to our results, we find that SSs can explain well the high-frequency
QPOs in the GFs of some SGRs. We further discuss the effects of magnetic field
on the torsional oscillation frequencies. The typical value of the magnetic
field strength is adopted as $B_{\mu} = 4 \times 10^{16}$\,G, which is much
larger than the ordinary NS models \citep{Sotani:2006at}. 

To explain the observed low-frequency QPOs in the GFs of some SGRs, we consider
that SSs may have a thin surface ocean
with a density in the range of $10^{6} $--$10^{9}\,\rm g \, cm ^{-3}$. The depth is
considered to be in the range of $10 - 50 $\,m. We estimate the frequencies of
the ocean-crust interface modes and find that the interface modes can interpret
well the observed low-frequency QPOs in GFs for some SGRs.


\citet{Watts:2006hk} have also investigated the thin nuclear crust model and the
quark nugget crust model. They calculated the frequencies of the torsional
oscillation modes using plane-parallel approximation and discussed the effects
of the magnetic field and the temperature on the frequencies of the torsional
modes. Compared with their results, when considering the magnetic field strength
$B_{\mu}$, the frequencies have increased by up to $25 \%$--$35 \%$, compared to
the frequencies of non-magnetized models. For the thin nuclear crust model, the
typical magnetic field value is $B_{\mu} \sim 4 \times 10^{13}$\,G. The
frequencies of the first overtone could increase up to $24\%$ at $B_{\mu} \sim 4
\times 10^{13}$\,G. The nugget crust model has a wider range of frequencies due
to its uncertainty in the strange quark mass $m_{\rm s}$. Our results show that
both the thin nuclear crust model and the nugget crust model are difficult to
reproduce well the recorded QPO frequencies.

Analysis of the magnetar QPOs in GFs could enable us to look for a correct
interpretation of their origin and the physical nature of the oscillations
\citep{LIGOScientific:2009rrx, Kalmus:2009uk, LIGOScientific:2010jrd,
LIGOScientific:2022sts}. In the future, the strong couplings of magnetar
oscillations to GWs will provide an excellent opportunity to apply the
asteroseismological methods to compact star studies, and eventually uncover the
nature of astrophysical compact objects.

\section*{Acknowledgements}
We thank the anonymous referee for comments.
This work was supported by the National SKA Program of China (2020SKA0120300,
2020SKA0120100), the National Natural Science Foundation of China (11975027,
11991053, 12275234, 12342027), the Max Planck Partner Group Program funded by
the Max Planck Society, and the High-Performance Computing Platform of Peking
University.

\section*{Data Availability}

The data underlying this paper will be shared on reasonable request to the
corresponding authors.



\bibliographystyle{mnras}
\bibliography{example} 


\bsp	
\label{lastpage}
\end{document}